# Quantum Laplacian Eigenmap

Yiming Huang[1], Xiaoyu Li[1]

[1] School of information and software engineering

University of Electronic Science and Technology of China, Chengdu, China

Laplacian eigenmap algorithm is a typical nonlinear model for dimensionality reduction in classical machine learning. We propose an efficient quantum Laplacian eigenmap algorithm to exponentially speed up the original counterparts. In our work, we demonstrate that the Hermitian chain product proposed in quantum linear discriminant analysis (arXiv:1510.00113,2015) can be applied to implement quantum Laplacian eigenmap algorithm. While classical Laplacian eigenmap algorithm requires polynomial time to solve the eigenvector problem, our algorithm is able to exponentially speed up nonlinear dimensionality reduction.

# Quantum Laplacian Eigenmap

Yiming Huang, Xiaoyu Li

School of information and software engineering

University of Electronic Science and Technology of China, Chengdu, China

**Abstract:** Laplacian eigenmap algorithm is a typical nonlinear model for dimensionality reduction in classical machine learning. We propose an efficient quantum Laplacian eigenmap algorithm to exponentially speed up the original counterparts. In our work, we demonstrate that the Hermitian chain product proposed in quantum linear discriminant analysis (arXiv:1510.00113,2015) can be applied to implement quantum Laplacian eigenmap algorithm. Compared with classical Laplacian eigenmap algorithm which requires polynomial time to solve dimensionality reduction, our algorithm is able to provide an exponential speedup.

Machine learning and data analysis play an increasingly key role, such as dimensionality reduction, prediction and classification. In many cases, the original data are always in extremely high-dimensional feature space. For instance, an image with $n^2$ pixels (that is, regarding each pixel as a feature) which yields $n \times n$ dimension feature space. So in order to analysis these original data with high-dimensional features, it is necessary to reduce dimension of it which has the natural structure of a low-dimensional manifold embedded in $\mathbb{C}^{n^2}$.

To analyze extremely large collection of data in high dimension, no matter what method of dimensionality reduction we choose, the time complexity should be considered. As we all know, well-designed quantum algorithms could provide us dramatically speed up classical algorithms. For example, Lloyd et al propose a quantum version PCA which exponentially speed up counterpart algorithm[1][2], and Cong et al generalized HHL algorithms to implement quantum discriminant analysis[3]. Nevertheless there is no work for nonlinear dimensionality reduction, therefore we propose a quantum Laplacian eigenmap algorithm (QLE), it applies the Hermitian chain product presented in [3] and matrix exponentiation technique presented in [4] to solve the nonlinear dimensionality reduction problem with the complexity $O(poly(\log(mn))/\varepsilon^3)$.

## 1. Classical Laplacian eigenmap algorithm

Laplacian eigenmap algorithm assumes that the data lies on or around a low-dimensional manifold in a high-dimensional space[5], it builds a graph $G=(V,E)$ which reflects the manifold structure of data by using the local information of each data point, and regards the data as the vertex and the similarity of different neighbor data point as the edge of this graph[6]. The task of dimensionality reduction problem is to minimize the objective function $J(u)$.

$$J(Y)=\sum_{i,j}(y_i - y_j)^2 w_{ij} = 2Y^T LY \qquad (1)$$

Where $y_i$ is the low-dimensional presentation of data point $x_i$, and $w_{ij}$ denotes the weight between data point $x_i$ and data point $x_j$. $L$ represents the Laplacian matrix of graph $G=(V,E)$ of data set. The optimization problem $\min(2Y^T LY)$ can be transformed to a generalized eigenvalue problem.

$$Lv = \lambda Dv \qquad (2)$$

Where $D$ is a diagonal matrix, $D_{ii} = \sum_j W(i,j)$, and $W(i,j)$ denotes the weight between data point $x_i$ and data point $x_j$. The $d$ minimum nonzero eigenvalue of eigenvector $v$ can construct the low-dimensional representation $Y$.

## 2. Quantum Laplacian eigenmap algorithm

Quantum Laplacian eigenmap regards the Laplacian matrix $L$ as the covariance matrix of data set which corresponds to a density matrix, so the original eigenvalue problem is equivalent to $D^{-1}Lv = \lambda v$, where $L = I \cdot I^\dagger$, and $I$ is the incidence matrix of the graph $G=(V,E)$. The incidence matrix stores the relationship between the each node and its connected edges. The following procedure will describe how to implement quantum Laplacian eigenmap.

**Step 1:** In our work, we first assume that we can efficiently use quantum RAM[7] to construct the matrix $D$ and matrix $L$. Define the covariance $L = I \cdot I^\dagger$, where $I$ has columns $a_i$, in quantum form $L=\sum_i |a_i||a_i\rangle\langle i|$. By applying the quantum RAM, we are able to construct the states $|\varphi\rangle=\sum_i |a_i||a_i\rangle|i\rangle$ and $|\psi\rangle=\sum_i |d_i||d_i\rangle|i\rangle$, and then the density matrix of the second register of these two states are proportional to L and D.

**Step 2:** We demonstrate that the Hermitian chain product technique presented in [3] can be used to solve $D^{-1}Lv = \lambda v$. Since the $L$ and $D$ are Hermitian positive matrix, the eigenvalue problem can be transformed to $F \cdot F^\dagger u = \lambda u$ where $F = L^{1/2} D^{-1/2}$. By using Hermitian chain product, we can construct $F \cdot F^\dagger$ in $O(\log(mn) k^{3.5}/\varepsilon^3)$.

**Step 3:** There are two ways to reveal the eigenvalue and the correspond eigenvector.

**Way 1:** Applying the matrix exponentiation technique proposed in [4], and then do quantum phase estimation[8], we can obtain the state $\sum_i \lambda_i |\varphi_i\rangle\langle\varphi_i| \otimes |\lambda_i\rangle\langle\lambda_i|$, where $\lambda_i$ and $\varphi_i$ are the eigenvalue and eigenvector of $F \cdot F^\dagger$.

**Way 2:** we use phase estimation to obtain the superposition of the eigenvalue and eigenvector, and apply amplitude amplification to reveal the eigenvalue and eigenvector. First of all, we initialize the registers $|\varphi_0\rangle=|0\rangle|0\rangle$, then apply the quantum Fourier transform to the first register, and Hadamard transform to the second register; $|\varphi_1\rangle=(U_{QFT} \otimes H^{\otimes m})|\varphi_0\rangle$. Next, Applying the series controlled-unitary operations to the second register, we obtain the state $|\varphi_2\rangle=(CU^{2^{m-1}}...CU^{2^0})|\varphi_1\rangle$,

where $U = e^{i2\pi F \cdot F^{\dagger}}$. After applying the inverse quantum Fourier transform to the first register, we can get the state $|\varphi_3\rangle = \sum_k \alpha_k |\lambda_k\rangle |\phi_k\rangle$. Since our goal is to reveal the eigenvalue and eigenvector, so amplitude amplification can help us find the marked eigenvalue and eigenvector. The two operators in amplitude amplification are $U_{mark} = (I - 2|\psi_u\rangle\langle\psi_u|) \otimes I^{\otimes m}$ and $U_{flip} = 2|\varphi_3\rangle\langle\varphi_3| - I$, The unitary operator $U_{mark}$ is used to mark the certain eigenvalues, and the operator $U_{flip}$ is used to reflect the state. Combing with these two operators and applying to $|\varphi_3\rangle$ in $O(1/|\lambda_i|)$ times, we can reveal the eigenvalues and the eigenvectors.

**Step 4:** By using Hermitian chain product and applying $L^{1/2}$ to eigenvector $u$, we can get the eigenvector of the original problem of (2).

We propose a quantum version of Laplacian eigenmap algorithm, QLE. Similar to researches [3] and [4], we also assume that we can efficiently use quantum random access memory to construct $L$ and $D$. While the classical Laplacian eigenmap algorithm requires polynomial time to solve the eigenvector problem, our algorithm can provides exponential speedup of it by using Hermitian chain product technique, which can perform to do nonlinear dimensionality reduction with $O(poly(\log(mn))/\varepsilon^3)$.